\documentclass[11pt,letterpaper]{article}

\usepackage[english]{babel}
\usepackage{soul} % handling hyphenation
\usepackage[normalem]{ulem}  % underlining
\usepackage{setspace}  % control spacing between lines
\usepackage{xcolor}  % advanced colors, e.g., custom colors
\usepackage{color}  % color management
\usepackage{setspace}  % change line spacing
\usepackage[compact,md]{titlesec}  % customize section title formatting
\usepackage{textcomp}  % support for special text symbols, such as bullets
\usepackage[version=4]{mhchem}  % typesetting chemistry notation
\usepackage{siunitx}  % SI units

\usepackage[utf8]{inputenc}  % allows for unicode (UTF8) not just ASCII 
\usepackage{authblk}  % author and affiliation
\usepackage{blindtext}  % creates 'blind' content for testing
\usepackage{url}  % verbatim inclusion of URLs
\usepackage{etoolbox}  % advanced programming tools for classes/packages
\usepackage{tcolorbox}  % Adding colored box with text and equations
\usepackage[export]{adjustbox}  % adjust boxed content
\usepackage{subfiles}  % enable subfiles to be able to run standalone

\usepackage{graphicx}  % including graphics
\usepackage{pdfpages}  % including external pdfs
\usepackage{pgfgantt}  % Gantt charts
\usepackage{caption}  % advanced captions
\usepackage{subcaption}  % captions for subfigures
\usepackage{rotating}  % rotate floats
\usepackage{wrapfig}  % enable wrapping text around figures/tables
\usepackage{tabularx}  % tables with adjustable-width columns
\usepackage{multirow}  % tables with cells spanning multiple rows
\usepackage{array}  % advanced arrays and tables
\usepackage{longtable}  % allow multi-page tables
\usepackage{makecell}  % advanced options for table layouts
\usepackage{colortbl}  % color for tables

\usepackage{amsmath}  % features for mathematical typesetting
\usepackage{amsfonts}  % extended fonts for mathematics
\usepackage{amssymb}  % extended symbols
\usepackage{amsthm}   % for theorems, proofs, lemmas, ...
\usepackage{esdiff}  % derivatives
\usepackage{siunitx}  % SI units
\usepackage{cancel}  % lines through math
\usepackage{bm}  % bold symbols in math

\usepackage{doi}  % create correct hyperlinks for DOIs
\usepackage{hypernat} % makes hyperref anf natbib to work together
\usepackage[sorting=none,backend=biber,maxnames=50,citestyle=numeric-comp]{biblatex}

\usepackage{algorithm}  % floating algorithm environment
\usepackage{algpseudocode}  % actual algorithms
\usepackage{listings}  % typesetting code

\titlespacing{\section}{0pt}{1.5ex}{0.5ex}
\titlespacing{\subsection}{0pt}{1ex}{0.5ex}
\titlespacing{\subsubsection}{0pt}{0.5ex}{0ex}

\titleformat{\section}[block]
{\center\normalfont\scshape}  % general formatting
{\thesection}  % the label and number
{0.5em}  % space between label/number and section name
{}  % formatting applied just to section name
[]  % punctuation or other commands following section name

\titleformat{\subsection}[block]
{\normalfont\bfseries}  % general formatting
{\thesubsection}  % the label and number
{0.5em}  % space between label/number and section name
{}  % formatting applied just to section name
[]  % punctuation or other commands following section name

\titleformat{\subsubsection}[block]
{\normalfont\itshape}  % general formatting
{\thesubsubsection}  % the label and number
{0.5em}  % space between label/number and section name
{}  % formatting applied just to section name
[]  % punctuation or other commands following section name

\newcommand\preprintheader[4]{
    \begin{tcolorbox}[boxrule=0.75pt, arc=2pt, coltext=ucsd_blue, colback=white,colframe=ucsd_gold, fontupper=\rmfamily]
        \footnotesize
        \setstretch{1.25}
        {\textcolor{ucsd_gray}
        {This is the preprint version of the following article:}}

        #1

        \vspace{1mm}

        \textcolor{ucsd_gray}{Published article:} \;
        \url{https://doi.org/#2}

        \textcolor{ucsd_gray}
        {Preprint pdf:} \;
        \url{#3}

        \textcolor{ucsd_gray}
        {Bibtex:} \;
        \url{#4}

    \end{tcolorbox}
    
    \vspace{-6mm}
}
\definecolor{ucsd_blue}{RGB}{24, 43, 73}
\definecolor{ucsd_gold}{RGB}{198, 146, 20}
\definecolor{ucsd_blue_light}{RGB}{0, 98, 155}
\definecolor{ucsd_gold_light}{RGB}{255, 205, 0}
\definecolor{ucsd_gray}{RGB}{116, 118, 120}
\newcommand\p\partial  % symbol in partial derivative
\newcommand\f\frac  % fraction

\usepackage{pythonhighlight}
\usepackage{setspace}

\graphicspath{{./Plots/}{./Figures/}}

\begin{document}

\title{\textbf{Accelerating model evaluations in uncertainty propagation on tensor grids using \\ computational graph transformations}
}
\author{Bingran Wang\footnote{Ph.D Candidate, Department of Mechanical and Aerospace Engineering.}, Mark Sperry\footnote{Ph.D Student, Department of Mechanical and Aerospace Engineering.}, Victor E. Gandarillas\footnote{Ph.D Candidate, Department of Mechanical and Aerospace Engineering.}, and John T. Hwang\footnote{Assistant Professor, Department of Mechanical and Aerospace Engineering.}}
\date{}
\affil{University of California San Diego, La Jolla, CA, 92093}

\renewcommand\Affilfont{\itshape\small}
\begin{refsection}
    \preprintheader
    {
        Bingran Wang, Mark Sperry, Victor E. Gandarillas, and John T. Hwang. Accelerating model evaluations in uncertainty propagation on tensor grids using computational graph transformations. Aerospace Science and Technology, 2024.
    }
    {10.1016/j.ast.2023.108843}
    {https://github.com/LSDOlab/lsdo_bib/blob/main/pdf/wang2024accelerating.pdf}
    {https://github.com/LSDOlab/lsdo_bib/blob/main/bib/wang2024accelerating.bib}
    
    {\let\newpage\relax\maketitle}
    \vspace{-6mm}
    % \setstretch{2}
    
 \begin{abstract}
     Methods such as non-intrusive polynomial chaos (NIPC), and stochastic collocation are frequently used for uncertainty propagation problems. Particularly for low-dimensional problems, these methods often use a tensor-product grid for sampling the space of uncertain inputs.
    A limitation of this approach is that it encounters a significant challenge: the number of sample points grows exponentially with the increase of uncertain inputs. Current strategies to mitigate computational costs abandon the tensor structure of sampling points, with the aim of reducing their overall count.
    Contrastingly, our investigation reveals that preserving the tensor structure of sample points can offer distinct advantages in specific scenarios. Notably, by manipulating the computational graph of the targeted model, it is feasible to avoid redundant evaluations at the operation level to significantly reduce the model evaluation cost on tensor-grid inputs. This paper presents a pioneering method: Accelerated Model Evaluations on Tensor grids using Computational graph transformations (AMTC). The core premise of AMTC lies in the strategic modification of the computational graph of the target model to algorithmically remove the repeated evaluations on the operation level.
    We implemented the AMTC method within the compiler of a new modeling language called the Computational System Design Language (CSDL).
    We demonstrate the effectiveness of AMTC by using it with the full-grid NIPC method to solve four low-dimensional UQ problems involving an analytical piston model, a multidisciplinary unmanned aerial vehicle design model, a multi-point air taxi mission analysis model, and a single-disciplinary rotor model, respectively. For three of the four test problems, AMTC reduces the model evaluation cost by between 50\% and 90\%, making the full-grid NIPC the most efficacious method to use among the UQ methods implemented. 
\end{abstract}
    \section{Introduction}
In practical scenarios, the numerical models inherently fall short of achieving perfect fidelity with reality and consistently contend with inherent uncertainties.
To address this challenge, uncertainty propagation, also known as forward uncertainty quantification (UQ), aims to undertake a rigorous examination and analysis of how a system's output is influenced by the uncertainties inherent in its inputs.
The resultant assessments of uncertainty serve as a valuable foundation for facilitating informed decision-making and thorough risk assessment.
The applications of UQ can be found in many scientific and engineering domains such as 
weather forecast~\cite{joslyn2010communicating,hess-9-381-2005}, structural analysis~\cite{wan2014analytical,hu2018uncertainty} and  aircraft design~\cite{ng2016monte, wang2023high,lim2022uncertainty}.
 
Uncertainties within the inputs can stem from diverse origins and are typically classified into two categories: aleatoric and epistemic. Aleatoric uncertainties, recognized as irreducible uncertainties, are intrinsic to the system and encompass factors such as fluctuations in operational conditions, mission prerequisites, and model parameters. These uncertainties can often be described within a probabilistic framework. 
In contrast, epistemic uncertainties constitute reducible uncertainties arising from knowledge gaps. These uncertainties may be induced by approximations utilized in computational models or numerical solution methods, and are often difficult to be described within a probabilistic framework. 
This paper focuses on UQ problems within a probabilistic formalism where uncertain inputs are continuous random variables with known probability density distributions. The goal is to estimate statistical moments such as the mean and variance of the Quantity of Interest (QoI), or more complex risk measures such as the probability of failure or the conditional value at risk.

For this type of UQ problem, the most common non-intrusive methods include the method of moments, polynomial chaos, kriging, and Monte Carlo. 
The method of moments analytically computes the statistical moments of the QoI, using the Taylor series approximation. It often only uses function evaluations and derivative information at one or a small number of input points~\cite{wooldridge2001applications}. 
The most commonly employed variants of this method utilize derivatives up to either first~\cite{fragkos2019pfosm} or second order~\cite{luo2018statistical,luo2020optimal} to ensure cost-effectiveness.
While adept at efficiently estimating statistical moments, this approach grapples with challenges in estimating other kinds of risk measures. Moreover, it may lack accuracy when input variance is significant.
In contrast, the Monte Carlo approach employs random sampling of uncertain inputs to compute the risk measures. Based on the law of large numbers, Monte Carlo's convergence rate is independent of the number of uncertain inputs, making it particularly attractive for solving high-dimensional problems. Recent strides in enhancing its efficiency encompass multi-fidelity~\cite{peherstorfer2016optimal,peherstorfer2018survey} and importance sampling~\cite{tabandeh2022review} techniques. However, in the context of low-dimensional problems, the Monte Carlo method might demand substantially more model evaluations to achieve the same level of accuracy as the alternative UQ methodologies.
Kriging, also known as Gaussian process regression, is a common statistical technique employed in various fields. In UQ, kriging constructs a surrogate response surface by leveraging input-output data pairs. The surrogate model replaces the original expensive function to allow a large number of model evaluations in order to perform reliability analysis~\cite{kaymaz2005application,hu2016single} or optimization under uncertainty~\cite{rumpfkeil2013optimizations}.
For low-dimensional UQ problems, if the objective function is assumed to be smooth, the polynomial chaos-based techniques often stand out as the most efficacious option.
The generalized polynomial chaos theory represents the QoI as orthogonal polynomials that are derived from the distributions of the uncertain inputs. Benefiting from the inherent smoothness of the random space, polynomial chaos-based methods exhibit rapid convergence rates facilitated by sampling or integration techniques.
Common polynomial chaos-based methods include non-intrusive polynomial chaos (NIPC)~\cite{hosder2006non,jones2013nonlinear,keshavarzzadeh2017topology} and stochastic collocation (SC)~\cite{xiu2005high, babuvska2007stochastic}.

For both NIPC and SC methods, we can use a tensor-product grid to sample the input space, but this approach is often not preferred and is only used for very low-dimensional UQ problems, as the number of input points increases exponentially with the number of uncertain inputs. The commonly used methods choose to use a sparse grid to sample the input space or have the sample points randomly generated, in order to reduce the input points with a minimum loss of accuracy.
However, our recent findings show that maintaining a tensor structure for the input points offers some unique advantages and for certain problems, the total model evaluation cost can be reduced in a different approach without changing the tensor structure of the input points.

When we evaluate a computational model on full-grid input points, 
if we view the computational model as a computational graph with only elementary operations, the current framework evaluates each operation the same number of times at the tensor-product input points of all the uncertain inputs. 
However, for each operation, the output data only has distinct values at the distinct input points in their dependent input space, and a large portion of the operations may not be dependent on all of the uncertain inputs.
This means the current framework of NIPC and SC could create many wasteful evaluations on a large portion of the operations.

In this paper, we propose a computational graph transformation method called Accelerated Model evaluations on Tensor grids using Computational graph transformations (AMTC) that algorithmically modifies the computational graph to eliminate the unnecessary evaluations incurred by the current framework of integration-based NIPC and SC.
AMTC reduces the model evaluation cost on full-grid quadrature points by partitioning the computational graph into sub-graphs using the operations' dependency information and evaluating each sub-graph on the distinct quadrature points.
The proposed method is implemented in conjunction with a new algebraic modeling language, the Computational System Design Language~(CSDL). 
We achieve cost reductions on UQ problems in a general and automatic way by analyzing and manipulating the computational graph that CSDL makes available.

This method has been applied to four different UQ problems with different computational graph structures. 
For three of the problems, we observed a significant acceleration in model evaluation time when solving the UQ problems using the full-grid NIPC method, which makes this UQ method the most efficient method among the implemented UQ methods for these problems.
Generally, for a wide range of UQ problems that involve multi-point, multi-disciplinary, or other models that possess a sparse graph structure, we expect our method to significantly reduce the model evaluation time on tensor-grid input points, which can improve the time scaling for, but not limited to NIPC and SC methods.

This paper is organized as follows. 
Section \ref{Sec: Background} gives some background on PCE-related UQ methods, the current model evaluation framework, and how to view tensor-grid evaluations through computational graphs. 
Section \ref{Sec: Methodology} presents the details of the new computational graph transformation algorithm, AMTC, and how it is implemented in CSDL. 
Section \ref{Sec: Numerical Results} shows the numerical results of three UQ problems.
Section \ref{Sec: Conclusion} summarizes the work and offers concluding thoughts.

    \section{Background}
\label{Sec: Background}

\subsection{Polynomial chaos expansion}
\iffalse
\textcolor{blue}{
\begin{itemize}
    \item PCE theory. Generalized polynomial chaos.
    \item Basis selection. Orthogonal Property. Extract statistical moments.
    \item NIPC method, regression and integration (full-grid, sparse grid)
    \item Stochastic collocation. How interpolation works. 
    \item Comparison between NIPC and SC. 
    \item Curse of dimensionality. NIPC and SC could require model evaluations result on tensor-product of quadrature points. The evaluation time could increase exponentially as the number of input increases.
\end{itemize}}
\fi

Wiener initially introduced the concept of polynomial chaos expansion (PCE), utilizing Hermite polynomials to model the response of a system affected by Gaussian uncertain inputs \cite{wiener1938homogeneous}. This foundational concept was later advanced into the generalized polynomial chaos (gPC) method by Xiu and Karniadakis \cite{xiu2002wiener}. Within the framework of gPC, model responses are represented through polynomial series of uncertain inputs. These polynomials are meticulously chosen to exhibit orthogonality with respect to the probability distribution of the uncertain inputs, thereby ensuring exponential convergence.

We address a general problem in uncertainty quantification (UQ) involving a function defined as follows:
\begin{equation}
f = \mathcal{F}(u),
\end{equation}
where $\mathcal{F}:\mathbb{R}^{d} \to \mathbb{R}$ represents the model evaluation function, $u \in \mathbb{R}^d$ denotes the input vector, and $f \in \mathbb{R}$ represents a scalar output. The uncertainties associated with the inputs are expressed as a stochastic vector denoted by $U := [U_1, \ldots, U_d]$, under the assumption that these random variables are mutually independent. The stochastic input vector adheres to the probability distribution $\rho(u)$ with its support defined by $\Gamma$. Our focus centers on the evaluation of risk measures of the output random variable, $f(U)$.

In gPC, the stochastic output, $f(U)$, is defined as an infinite series of orthogonal polynomials with respect to the uncertain input variables:
\begin{equation}
\label{eqn: PCE}
    f(U) = \sum_{i = 0}^{\infty} \alpha_i \Phi_i(U),
\end{equation}
where $\Phi_i(U)$ are the PCE basis functions that are generated based on $\rho(U)$, and $\alpha_i$ are the corresponding weights that need to be determined.

Choosing the PCE basis functions is important to ensure the rapid convergence of PCE-based methods. These PCE basis functions have to satisfy the orthogonality property,
\begin{equation}
\label{Eqn: orthogonal property}
    \left< \Phi_i(U), \Phi_j(U) \right> = \delta_{ij},
\end{equation}
where $\delta_{ij}$ is Kronecker delta and the inner product is defined as
\begin{equation}
    \left< \Phi_i(U), \Phi_j(U) \right> = \int_{\Gamma}  \Phi_i(u) \Phi_j(u)\rho(u) du.
\end{equation}

In one-dimensional problems, we can directly use the univariate orthogonal polynomials as PCE basis functions. In Table~\ref{tab: orthog poly}, we show the univariate orthogonal polynomials for common types of continuous random variables. In multi-dimensional problems, if the input variables are mutually independent, the PCE basis functions can be formed as a tensor-product of the univariate orthogonal polynomials corresponding to each uncertain input. To elucidate this, we introduce the concept of multi-index notation and define a $p$-tuple as
\begin{equation}
i^{\prime} = (i_1,\ldots, i_p) \in \mathbb{N}^p_0,
\end{equation}
with its magnitude determined as
\begin{equation}
|i^{\prime}| = i_1 + \ldots + i_p.
\end{equation}
Employing this notation, we denote the collection of univariate PCE basis functions for variable $u_k$ as $\{ \phi_{i}(u_k) \}_{i = 0}^p$. Then, the multivariate PCE basis with an upper limit on the total degree of $p$ can be expressed as
\begin{equation}
\label{eqn: multivariate}
    \Phi_{\boldsymbol{i^\prime}}(u) = \phi_{i_1}(u_1)\ldots \phi_{i_d}(u_d), \quad |i^\prime| \leq p. 
\end{equation}
In practice, one may use the PCE basis functions in \eqref{eqn: multivariate} to truncate the infinite series in \eqref{eqn: PCE}, resulting in:
\begin{equation}
\label{Eqn: pce estimate}
    f(U) \approx \sum_{i = 0}^{q} \alpha_i \Phi_i(U).    
\end{equation}
The resultant number of PCE basis functions, $q + 1$, satisfies:
\begin{equation}
    q + 1 = \frac{(d + p)!}{d
    ! p!}.
\end{equation}
Once the PCE coefficients are estimated, it becomes straightforward to compute statistical moments such as the mean and standard deviation of the model output. These calculations can be expressed as follows:
\begin{equation}
   \mu_f= \alpha_0,
\end{equation}
\begin{equation}
   \sigma_f = \sum_{i = 1}^d\alpha_i^2 .
\end{equation}
For risk measures like the probability of failure, the truncated PCE model can be treated as a surrogate model, and methods like Monte Carlo can be applied on the surrogate model to compute the desired risk measure.
\begin{table}[]
\caption{Orthogonal polynomials for common types of continuous random variables}
\centering
\begin{tabular}{c c c c} 
 \hline
 Distribution & Orthogonal polynomials & Support range \\ 
 \hline\hline
 Normal & Hermite  & $(-\infty, \infty)$ \\
 \hline
 Uniform & Legendre  & $[-1, 1]$  \\
 \hline
 Exponential & Laguerre  & $[0, \infty)$ \\
 \hline
 Beta & Jacobi  & $(-1, 1)$ \\
 \hline
 Gamma & Generalized Laguerre  & $[0, \infty)$ \\
 \hline
\end{tabular}
\label{tab: orthog poly} 
\end{table}

\subsubsection{Non-intrusive polynomial chaos method}
The non-intrusive polynomial chaos (NIPC) method solves the PCE coefficients $\alpha_i$ in \eqref{Eqn: pce estimate} by either integration or regression.
The integration approach uses the orthogonal property in \eqref{Eqn: orthogonal property} and projects the model output onto each basis function: 
\begin{equation}
    \alpha_i = \frac{\left< f(U), \Phi_i \right> }{\left< \Phi_i^2 \right>} = \frac{1}{\left< \Phi_i^2 \right>} \int_u f(u) \Phi_i(u) \rho(u) du.
\end{equation}
This requires solving a multi-dimensional integration problem, which is often estimated by using the Gauss quadrature approach:
\begin{equation}
\begin{aligned}
    \alpha_i & = \frac{1}{\left< \Phi_i^2 \right>}  \int_{u_1}\ldots \int_{u_d}  f(u_1, \ldots, u_d) \Phi_k(u_1,\ldots, u_d)\rho(u_1,\ldots, u_d) du_1\ldots du_d \\
    & \approx \frac{1}{\left< \Phi_i^2 \right>} \sum_{i_1 = 0}^{k} \ldots \sum_{i_d = 0}^{k} w_1^{(i_1)}\ldots w_d^{(i_d)} f(u_1^{(i_1)}, \ldots ,u_d^{(i_d)})\Phi_i(u_1^{(i_1)}, \ldots ,u_d^{(i_d)}), 
\end{aligned}
\end{equation}
where $(u_i^{(1)},\ldots, u_i^{(k)})$ and $(w_i^{(1)},\ldots, w_i^{(k)})$ are the nodes and weights for the 1D Gauss quadrature rule in the $u_i$ dimension, $k$ is the number of quadrature points used in each dimension. 
The 1D quadrature rule with $k$ nodes is chosen such that the quadrature rule can exactly integrate 1D polynomials up to $(2k-1)$th order. In the  multi-dimensional case, the Gauss quadrature rule forms a tensor product of the 1D quadrature rule, which requires us to evaluate the model output at the tensor-product quadrature points, defined as
\begin{equation}
\label{eqn: full tensor}
 \boldsymbol{u} = \boldsymbol{u}_1^k \times \cdots \times \boldsymbol{u}_d^k,
\end{equation}
where 
\begin{equation}
\boldsymbol{u}_i^k:= \{u_i^{(j)}\}_{j = 1}^k.
\end{equation} 
The total number of quadrature points is thus $k^d$, and this number increases exponentially with respect to the number of uncertain input variables.
For high-dimensional problems, one way to mitigate the exponential growth of quadrature points is to use the Smolyak sparse grid approach ~\cite{nobile2008sparse}, which drops the higher-order cross terms in the quadrature points with minimal loss of accuracy. The sparse grid quadrature points can be expressed as:
\begin{equation}
\label{eqn：sparse grid}
    \boldsymbol{u} = \bigcup_{\ell-d+1 \leq|i| \leq \ell}\left(\boldsymbol{u}^{i_{1}}_1 \times \cdots \times \boldsymbol{u}^{i_{d}}_d\right),
\end{equation}
where $\ell$ is the level of construction. The existing variations of sparse-grid methods can be found in \cite{gerstner1998numerical}.
Another way to reduce the total number of quadrature points is to use a designed quadrature approach. 
In \cite{keshavarzzadeh2018numerical}, Keshavarzzadeh et al. formulate an optimization problem to optimize for the designed quadrature points and the corresponding weights that ensure the exact integration on a polynomial subspace. The number of designed quadrature points can be significantly smaller than the full-grid quadrature points for solving the high-dimensional integration problems for the same level of accuracy.

The regression approach first generates $n$ sample points for $U$ and evaluates the model for each sample point. 
The coefficients are determined by solving a linear least-squares problem:
\begin{equation}
    \begin{bmatrix}
    \Phi_1(u^{(1)}) &  \ldots & \Phi_q(u^{(1)}) \\
    \vdots & & \vdots \\
    \Phi_1(u^{(n)}) &  \ldots & \Phi_q(u^{(n)}) \\
    \end{bmatrix}
    \begin{bmatrix}
    \alpha_1 \\
    \vdots \\
    \alpha_q 
    \end{bmatrix} =
    \begin{bmatrix}
    f(u^{(1)}) \\
    \vdots \\
    f(u^{(n)})
    \end{bmatrix}.
\end{equation}
The rule of thumb is that the number of samples $n$ needs to be 2-3 times the number of coefficients $q$. This means the cost of the regression-based NIPC method can be reduced by using the sparse PCE basis, resulting in fewer coefficients to be estimated. Recent advances focus on adaptive basis selection and adaptive sampling\cite{blatman2008sparse,thapa2020adaptive,luthen2021sparse}.

\subsection{Model evaluation cost for current non-intrusive UQ methods}
If we consider only the non-adaptive approaches, the current framework for non-intrusive UQ methods includes three steps:

\begin{enumerate}
    \item Generate sample/quadrature points based on the distribution of uncertain inputs.
    \item Evaluate the target model for each of the sample points.
    \item Post-process the output points to compute the desired quantities of the output.
\end{enumerate}
Step one generates the input points based on the distribution of uncertain inputs. 
For methods like regression-based NIPC, Monte Carlo and kriging, the input points are often generated randomly (e.g. random sampling, Latin hypercube sampling). For integration-based NIPC and SC methods, they can be generated using the full/sparse-grid strategy. 
In step two, the target model is repeatedly evaluated at each of the input points.
For some models that support vectorized input, the input points can be vectorized and evaluated at the same time.
However, most of the time, the target model needs to be run in a for-loop by evaluating one input point at a time.
In step three, the model output points are used to estimate the PCE coefficients for NIPC, form Lagrange interpolation functions for SC, or construct Gaussian process model for kriging, before computing the desired quantity for the output. 

Under this framework, the model evaluation cost for any UQ method can be approximated as:
\begin{equation}
\label{eqn: linear cost}
    \text{cost} \approx n O(\mathcal{F}(u)),
\end{equation}
where $n$ is the number of sample points and $O(\mathcal{F}(u))$ is the model evaluation cost for one sample point.
When we use the full-grid approach for integration-based NIPC or SC to generate the input points, 
the number of input points increases exponentially with the dimension of uncertainty inputs as $n =  k^d$. Under the current framework to evaluate the model, the evaluation cost follows Eq. \ref{eqn: linear cost} and thus increases exponentially as the number of uncertain inputs, thus suffers the curse of dimensionality.

The current methods (e.g. sparse-grid and designed quadrature) to overcome the curse of dimensionality all focus on reducing the number of input points, $n$, by breaking the full-tensor structure of the input points, so that the input points do not increase exponentially as the number of the uncertain inputs, but none tries to break the linear relationship in Eq. \ref{eqn: linear cost} to reduce the evaluation cost on tensor-grid input points by using a computational graph transformation approach.
\subsection{Viewing tensor-grid evaluations via computational graph}
\begin{figure}[hbt!]
\centering
  \includegraphics[width=4cm]{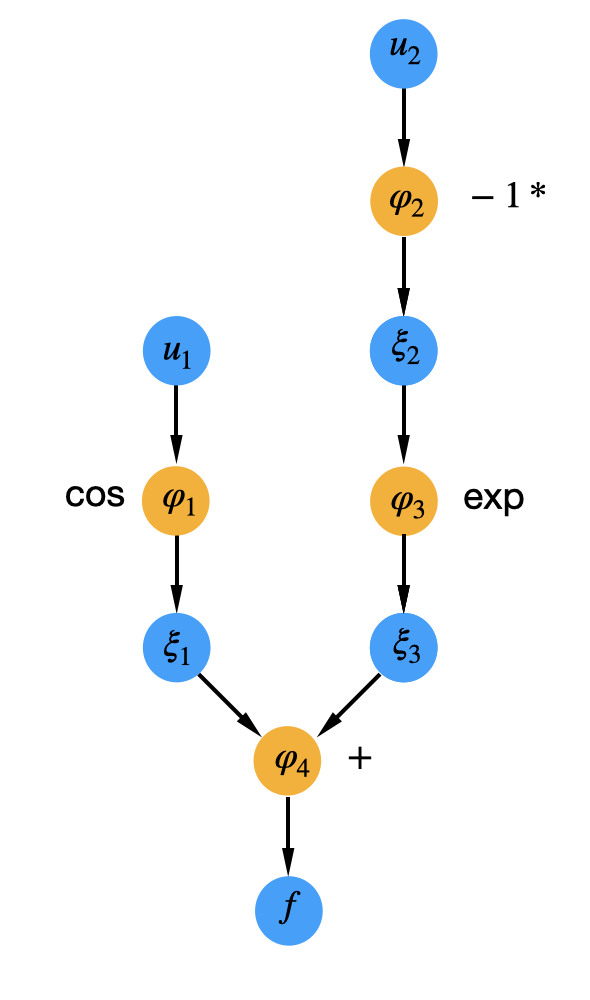}
\caption{Computational graph of function $f = cos(u_1) + exp(-u_2)$}
\label{fig:graph1}
\end{figure}

\iffalse
\textcolor{blue}{
\begin{itemize}
    \item Any computer program can be decomposed into elementary functions and operations.
    \item Graph describes computational process, directed acyclic graph. Operation with most two arguments.
    \item Most famous application is automatic differentiation. It uses graph to automate gradient computation.
    \item Tensorflow uses graph optimzier to speed up machine learning model evaluation costs. Reduce redundant nodes.
\end{itemize}}
\fi
Any computer program that describes a computational model is a computational process that executes a sequence of elementary functions and operations.
A computational process can be described by a computational graph. 
A computational graph is a directed acyclic graph in which the nodes represent the operations and variables, and the directed edges represent how operations and variables are connected to each other.
Such a graph is also bipartite because each operation is connected to an output variable, and likewise, variables are connected to operations (as arguments).
For example, the computer program that evaluates a function $f = cos(u_1) + exp(-u_2)$ can be decomposed into the following set of fundamental operations:
\begin{equation}
\label{eqn: computational process}
    \begin{aligned}
    & \xi_1 =  \varphi_1 (u_1) = cos (u_1); \\
    & \xi_2 = \varphi_2 (u_2) =  - u_2; \\
    & \xi_3 =  \varphi_3 (\xi_2) =  exp (\xi_2); \\
    & f = \varphi_4 (\xi_1, \xi_3) =  \xi_1 + \xi_3, \\
    \end{aligned}
\end{equation}
The computational graph for this model is shown in Fig.~\ref{fig:graph1} with the blue nodes representing the variables and the orange nodes representing the operations.
A common use of computational graphs is in automatic differentiation (AD)~\cite{baydin2018automatic,sperry2023automatic} to automatically compute the total derivatives. 
AD views the computer program as a sequence of operations and functions and repeatedly applies the chain rule to compute the total derivative. 
Another use of the computational graph is to modify the computational graph to reduce the time of memory cost of the model. 
In Tensorflow~\cite{abadi2016tensorflow}, a graph optimization method is implemented to reduce memory usage and evaluation costs for neural network models. It modifies the computational graph through constant folding, arithmetic simplification, and optimization.

If we view the model as a computational graph, in the current UQ framework, each operation and function is evaluated the same number of times as the number of input points. 
However, this could create many unnecessary evaluations when we evaluate the model on tensor-grid input points. For example, consider a UQ problem involving the simple function we showed before, given by
\begin{equation}
\label{eqn: simple func}
    f = \text{cos}(u_1) + \text{exp}(-u_2).
\end{equation}
In this problem, $U_1$ and $U_2$ are independent uncertain inputs, and we aim to compute the random variable $f(U_1,U_2)$. The computational process that describes this function is in \eqref{eqn: computational process}.
If we choose the full-grid quadrature points with $k$ quadrature points in each dimension, we need to evaluate this function at a total of $k^2$ quadrature points which are
\begin{equation}
\mathbf{u}  = \begin{Bmatrix}
(u_1^{(1)}, u_2^{(1)}) & \ldots & (u_1^{(k)}, u_2^{(1)}) \\
\vdots & \ddots & \vdots \\
(u_1^{(1)}, u_2^{(k)}) & \ldots & (u_1^{(k)}, u_2^{(k)}) \\
\end{Bmatrix}.
\end{equation}
For both inputs $u_1$ and $u_2$, if we view their input points as  vectors, they are in a tensor-product form between the quadrature points in its dimension and a vector of ones:
\begin{equation}
\label{Eqn: quadrature points vector}
\begin{aligned}
    u_1 = \text{vec}\left(\begin{bmatrix}
    u_1^{(1)} &\dots & u_1^{(k)} \\
    \vdots & \ddots & \vdots \\
    u_1^{(1)} & \ldots & u_1^{(k)}\\
    \end{bmatrix}  \right) = \text{vec}\left([u_1^{(1)}, \ldots u_1^{(k)}] \otimes [1, \ldots, 1] \right), \\
    u_2 = \text{vec}\left(\begin{bmatrix}
    u_2^{(1)} & \ldots & u_2^{(1)} \\
    \vdots & \ddots & \vdots \\
    u_2^{(k)} & \ldots & u_2^{(k)}\\
    \end{bmatrix}  \right) = \text{vec}\left( [1, \ldots, 1] \otimes [u_2^{(1)}, \ldots u_2^{(k)}]  \right).
    \end{aligned}
\end{equation}
Even though both vectors have a size of $k^2$, there are only $k$ distinct values, which are the values of the $k$ quadrature points in its dimension.
In the current framework, we can view the model evaluation step as propagating the input vectors in \eqref{Eqn: quadrature points vector} through the model's computational graph, which is shown in Fig. \ref{fig:graph1}. When we view the propagation of input vectors from an operational level, the current framework creates many repeated evaluations.
For example, when we evaluate the first operation node in the computational graph,  $\xi_1 = \text{cos}(u_1)$, we will evaluate this operation for $k^2$ points, since there are $k^2$ input points for $u_1$. 
However, out of the $k^2$ evaluations, there are at most $k$ distinct evaluations and thus the current framework creates $k^2 - k$ wasteful repeated evaluations in this case.
Similarly, for $\xi_2 = -1*u_2$ and $\xi_3 = \text{exp}(\xi_2)$, both of the operations' outputs are only dependent on $u_2$, and they will also have at most $k$ distinct values to be evaluated for. The current framework also creates $k^2 - k$ repeated evaluations for these operations. 
In contrast, for the last operation node, $f = \xi_1 + \xi_3$, there are $k^2$ distinct values to be evaluated as $f$ is dependent on both $u_1$ and $u_2$.

In practical UQ problems, we could have a large computational graph with a number of uncertain inputs, but some of the inputs may only affect a small part of the computational graph. For example, for a multidisciplinary model, we may have uncertain inputs coming from different disciplines. When we view the computational graph for the whole model, there may be a small portion of operations that are dependent on some uncertain inputs.
In this case,  we could reduce the total computational cost on the tensor-grid inputs by eliminating the repeated evaluation cost for each operation node but the output data still carry the necessary results to compute the results of the QoI.

The logic here is simple and straightforward. One can manually modify the size of the input and state variables in the model to achieve this reduction if the model has a simple graph structure.
However, for a complicated model with a large number of uncertainties, it becomes impractical to manually change the code to perform only necessary evaluations and match the shape of input for each node in the computational process. 

    \section{Methodology}
\label{Sec: Methodology}
\iffalse
\textcolor{blue}{
\begin{itemize}
    \item The number of necessary evaluations for each node is based on the number of input random variable it is related to.
    \item We generate this information from the graph, by tracing to the input for each node.
    \item Our logic her is to store each random variable as tensor, which contains its deterministic and random dimensions. deterministic case Rn, random dimensions Rkxn
    \item Pass into quadreature points only. It is only related to itself. Not the matrix.
    \item Univariate function have the consistent depency for input and output. Bivariate sometimes don't.
    \item Random dimension
    \item Increase the dimension of it using Einsum function to match the dimension of the tensors.
    \item We accomplish this by implementing a graph modification algorithm in CSDL.
\end{itemize}}
\fi
Here, we present the \textit{Accelerated Model evaluations on Tensor grids using Computational graph transformations} (AMTC) method. The goal of our method is to enable automatic generation of a modified computational graph, given a computer program, that eliminates all wasteful evaluations due to tensor-product sampling within a NIPC or SC algorithm.
The modified computational graph leverages the graph structure of the target model and enables performing the minimal number of evaluations for each operation.
To achieve this, we modify the size of input data that are passed into the operations.

In the UQ context, when we evaluate the computational model on full-grid quadrature points, we are forming a tensor product grid from the quadrature points in each uncertain input dimension. If we use $k$ quadrature points in each dimension, we end up having $k^d$ number of input points for $d$ uncertain inputs. For each dimension, these $k$ quadrature points are determined based on the probability distribution of that uncertain input.
When we view the target model monolithically, the output is dependent on all of the $d$ uncertain inputs. In the $d$-dimensional uncertain input space, we have a total of $k^d$ quadrature points, and it is required for us to evaluate the model output on all $k^d$ input points to gather all of the necessary information to determine the effect of the uncertain inputs on the model output for UQ purpose.
However, when we view it as a computational graph and consider the output of each operation in the graph, not all of the operations' outputs are dependent on $d$ uncertain inputs. For an output dependent on $d^{\prime}$ uncertain inputs with $d^{\prime} < d$, in its uncertain input space, there are only $k^{d^{\prime}}$ number of quadrature points, and accordingly this output only needs to be computed $k^{d^{\prime}}$ times to compute the necessary information for UQ purpose. 
Drawing from this rationale, we want to partition the model's computational graph into discrete sub-graphs. These sub-graphs are chosen such that the operations within each are dependent on the same uncertain inputs and thus need to be evaluated on the same quadrature points.
We achieve this by generating the dependency information for each operation, which shows whether the operation is dependent on an uncertain input. 
The dependency information is stored as an \textit{influence matrix}, written as $D(\varphi_i, u_j)$, whose  $(i,j)$th entry is 1 if the $i$th operation depends on the $j$th uncertain input and 0 otherwise. For example, the dependency information for the simple function in \eqref{eqn：sparse grid} is shown in Tab.\ref{tab: dep inf}.
\begin{table}[]
\caption{Operations dependency information for function $f = cos(u_1) + exp(-u_2)$}
\centering
\begin{tabular}{c | c c} 
 $D(\varphi_i, u_j)$ & $u_1$ & $u_2$ \\ 
 \hline
 $\varphi_1$ & 1 & 0 \\

 $\varphi_2$ & 0  & 1  \\

 $\varphi_3$ & 0  & 1 \\

 $\varphi_4$ & 1  & 1 \\
\end{tabular}
\label{tab: dep inf} 
\end{table}

From the dependency information, the operations that have the same dependency information on all of the uncertain inputs can be grouped together to form a sub-computational graph to be evaluated on the same quadrature points.
Each sub-computational graph should be evaluated the same number of times as the number of quadrature points in its uncertain input space. However, the output of one sub-graph may be the input of another sub-graph, since they have different uncertain input spaces, it is required to extend the uncertain input space of that output to match the uncertain input space of the second sub-graph's output. 
This requires us to perform a deliberate tensor-algebra operation. Fortunately, we can use the Einstein summation (Einsum) operation to achieve that. The Einsum operation can be used for compactly performing summations involving indices that appear in multiple tensors/matrices. In this case, we can treat the output of the sub-graph as a tensor and use the Einsum operation to perform a summation on specific indices between the output tensor and a tensor of ones, in order to extend its uncertain input space to a desired higher-order input space.
For example, consider an addition operation in the computational graph
\begin{equation*}
f = \xi_1 + \xi_2
\end{equation*}
with $\xi_1(u_1) \in \mathbb{R}^{k_{u_1}}$ and $\xi_2(u_2) \in \mathbb{R}^{k_{u_2}}$. 
In this case, $\xi_1$ is only dependent on $u_1$ with a size of $k_{u_1}$, while $\xi_2$ is only dependent on $u_2$ with a size of  $k_{u_2}$. The values in $\xi_1$ and $\xi_2$  correspond to their evaluations on the quadrature points in $u_1$ and $u_2$ dimensions, respectively.
The output of this operation, $f$ is dependent on both $u_1$ and $u_2$ and its vector should have a size of $\mathbb{R}^{k_{u_1}k_{u_2}}$. What we need to do here is to insert operations to modify the sizes of the inputs so that they have the same size as the output $f$ is supposed to be. Thus we evaluate the addition operation,  it results in the correct vector for it. 
We show a demonstration of how to extend the input size in a Python implementation using NumPy's Einstein summation function followed by a reshape operation. The specific code in Python would be
\begin{python}
import numpy as np
original_shape_xi_1 = (k_u1,1) #Original size for xi_1
original_shape_xi_2 = (k_u2,1) #Original size for xi_2
modified_shape = (k_u1*k_u2, 1) #Target size for xi_1 and xi_2
xi_1 = np.einsum('i...,p...->pi...', xi_1, np.ones((k_u2, 1)) #(k_u1,1) -> (k_u1,k_u2,1) 
xi_2 = np.einsum('i...,p...->ip...', xi_2, np.ones((k_u1, 1)) #(k_u2,1) -> (k_u1,k_u2,1) 
xi_1 = np.reshape(xi_1, modified_shape) #(k_u1,k_u2,1) ->(k_u1*k_u2, 1)
xi_2 = np.reshape(xi_2, modified_shape) #(k_u1,k_u2,1) ->(k_u1*k_u2, 1).
\end{python}
The code above transforms both inputs into the same size of the output with the correct order of values in the vectors.
In fact, for any dimension and size of the input, we can always use the Einstein summation functions followed by reshape functions to transform the input data to the correct size we need.
In our method, we insert the $einsum$ operation nodes for any connections between the sub-graphs we partitioned. Each $einsum$ operation node has an Einstein summation function followed by a reshape function with the right arguments to ensure the correct data flow between the sub-graphs.
The outline for the AMTC method is shown in Alg.\ref{alg: ATE}.
\begin{algorithm}
        \setstretch{1.17}
        \small
        \caption{Accelerated Model evaluations on Tensor grids using Computational graph transformations (AMTC) algorithm}
        \label{alg: ATE}
        \begin{algorithmic}[1]
        \State Specify a computational graph for a computational model 
        \State Run Kahn's algorithm\cite{kahn1962topological} to determine the correct order for the operations to be evaluated
        \State Generate the operations' dependency information and store it as an influence matrix
        \State Group the operations that share the same dependency information and form the sub-graphs 
        \State Insert the \textit{einsum} operation nodes with the right argument between the sub-graphs 
        \State Evaluate the modified computational graph with the correct input points
        \end{algorithmic}
\end{algorithm}

\begin{figure}%
    \centering
    \subfloat[\centering Computational graph without using AMTC]{{\includegraphics[width=5cm]{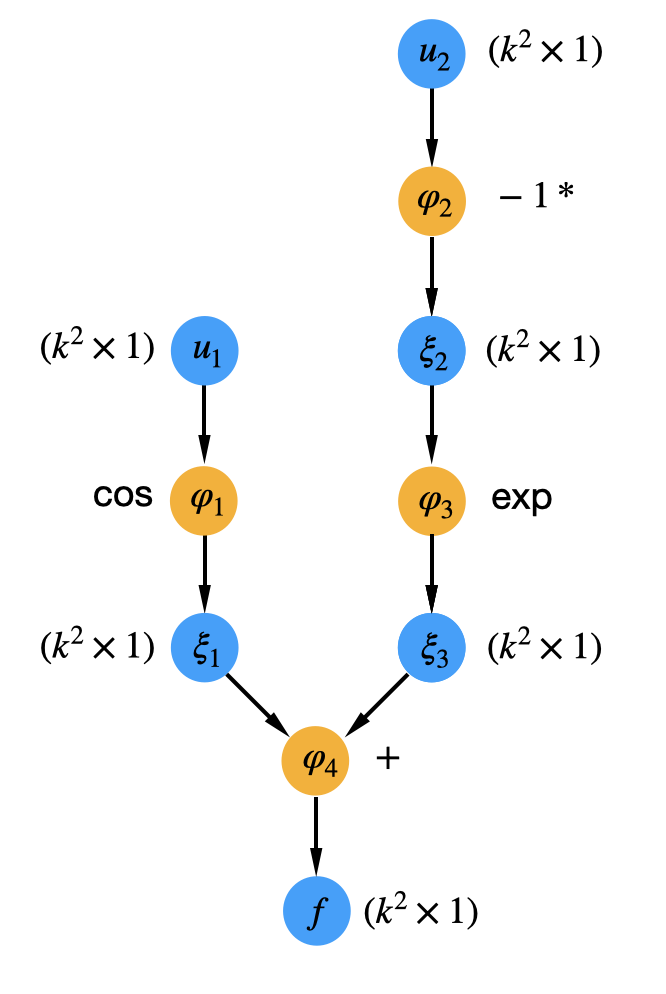} }}%
    \qquad
    \subfloat[\centering Computational graph using AMTC]{{\includegraphics[width=5cm]{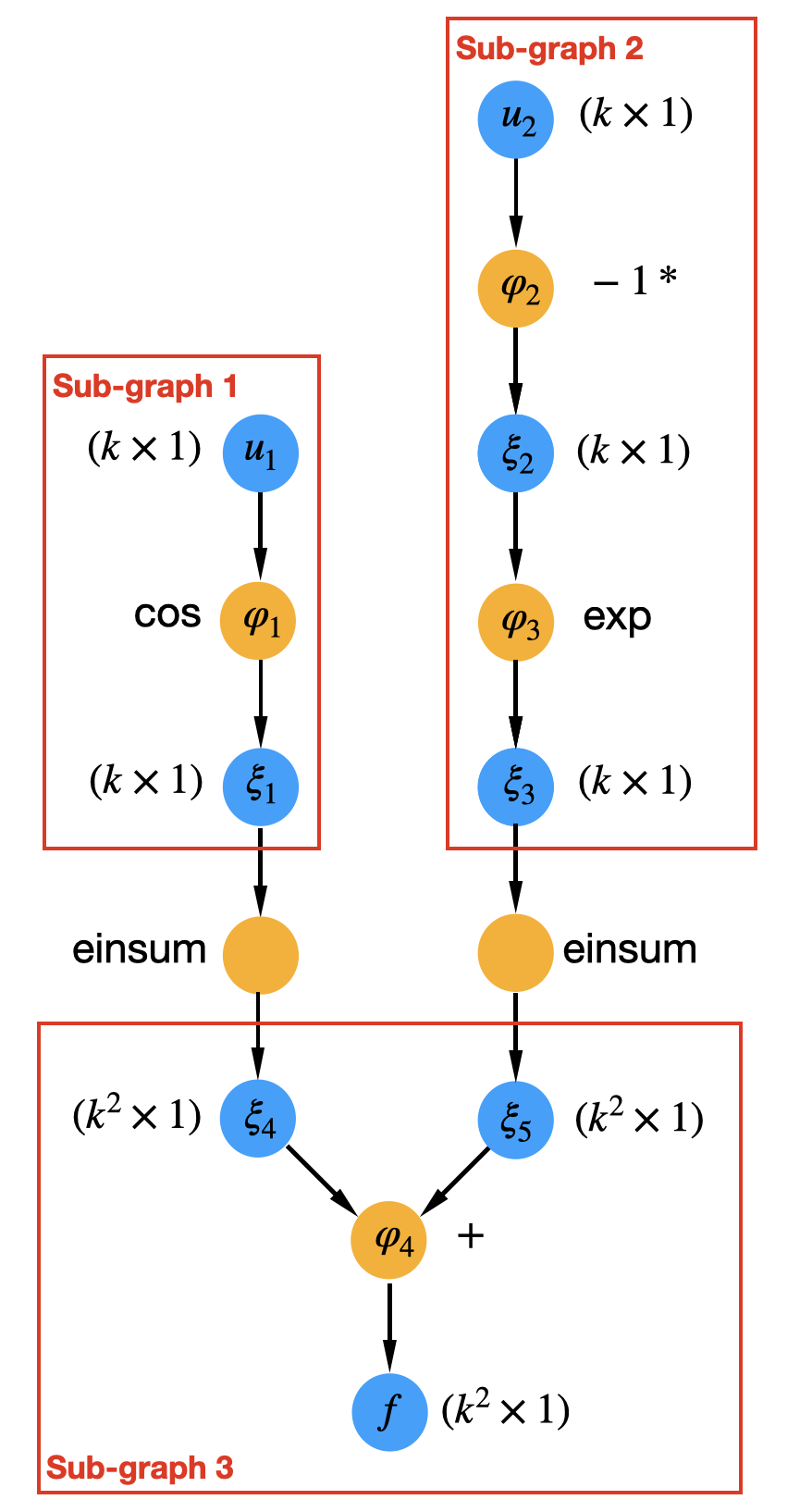} }}%
    \caption{Computational graphs with data size for full-grid quadrature points evaluation on $f = cos(u_1) + exp(-u_2)$}%
    \label{fig:graph comparison}%
\end{figure}

To demonstrate how our algorithm modifies the computational graph and reduces the evaluation costs, we show a comparison of computational graphs with and without using AMTC in Fig.~\ref{fig:graph comparison}.
Assuming we evaluate the function in \eqref{eqn: simple func} at full-grid quadrature points of the uncertain inputs, this figure shows the computational graphs with and without using the AMTC method.
The size of the data flow is also labeled in the graphs and the partitioned sub-graphs are indicated on the modified computational graph.
In this case, without using the AMTC method, both inputs $u_1$ and $u_2$ have the size of $k^2 \times 1$, and thus each operation in the computational graph is evaluated for $k^2$ number of times. However, with the AMTC method, the computational graph is partitioned into three sub-graphs. The operations in sub-graph 1 are evaluated on the $k$ quadrature points in $u_1$ dimension while  operations in sub-graph 2 are evaluated on the $k$ quadrature points in $u_2$ dimension. Two \textit{einsum} operations are inserted connecting sub-graph 1 and sub-graph 2 to sub-graph 3, so that the outputs of the first two sub-graphs are modified into the correct size and sub-graph 3 can be evaluated $k^2$ times to gather the same output values as the current method.
\subsection{Implementation in CSDL}
\iffalse
\textcolor{blue}{
\begin{itemize}
    \item CSDL has front-end, middle-end, back-end.
    \item User defines the physcis of the model in front end.
    \item Middle end (graph optimizer) generates the computational graph to pass into the back-end.
    \item Back-end executes the evaluations follow the computational graph.
    \item Our algorithm is implemented as an extra step in the middle end.
    \item It takes in the quadrature points as a vector and add the extra nodes of Einsum and reshape functions to the graph.
\end{itemize}}
\fi
The AMTC method can only be implemented from the data access layer of a software package where we have access to the computational graph and data structure of a computer program. 
This is possible thanks to the Computational System Design Language (CSDL) package\footnote{CSDL software repository: \url{https://lsdolab.github.io/csdl/}}~\cite{gandarillas2022novel}. 
CSDL is an embedded domain-specific language targeting large-scale multidisciplinary design analysis and optimization problems. 
With CSDL, the user defines the model as a sequence of operations by using a software interface that is highly expressive and natural (CSDL code resembles regular Python).

\begin{figure}[hbt!]
\centering
  \includegraphics[width= 16 cm]{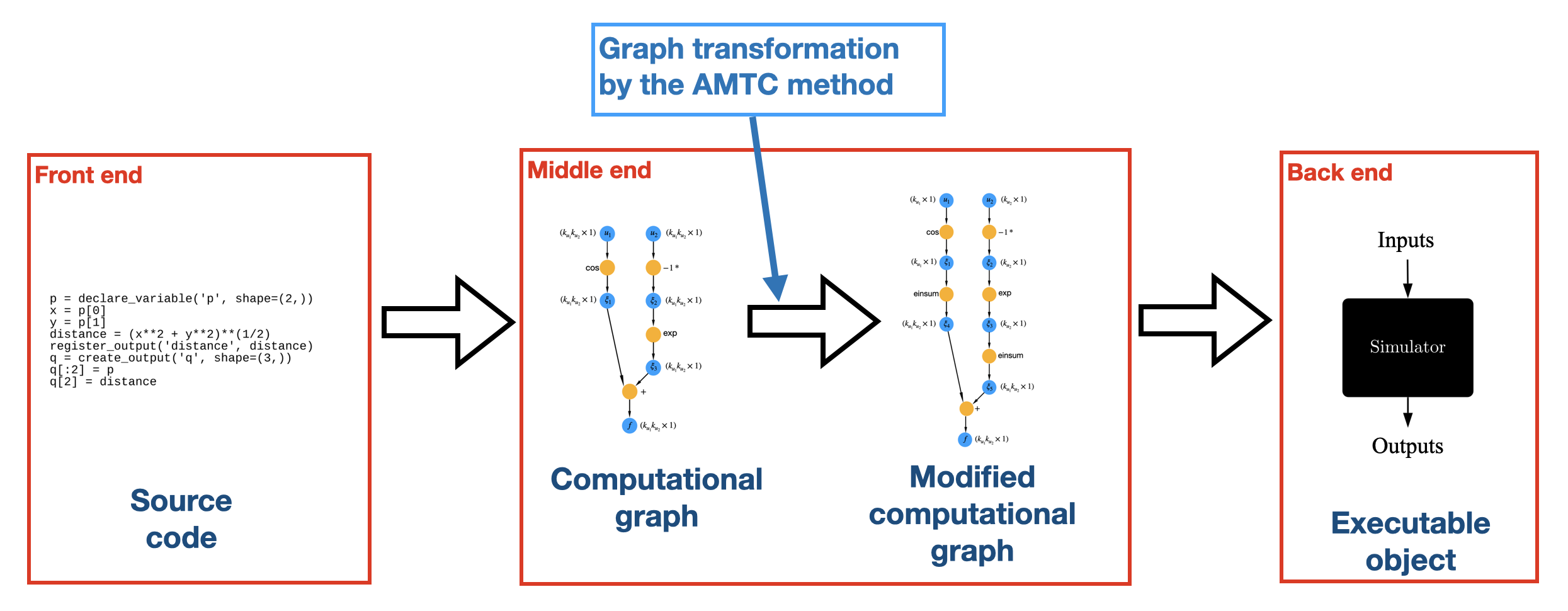}
\caption{Demonstration for the implementation of AMTC on CSDL}
\label{fig: ATE}
\end{figure}

We show a demonstration graph for the implementation of AMTC on CSDL in Fig.~\ref{fig: ATE}. The CSDL package has a unique three-stage compiler, which includes front-end, middle-end, and back-end. Using the front-end, the user defines the model and specifies the problem that needs to be solved. Next, the middle-end constructs a computational graph for that model, and the AMTC acts as a graph transformation method to generate the modified computational graph that is more efficient to evaluate. Finally, at the back-end, it creates an executable script based on the modified computational graph using an automatic code generation approach. 

% Therefore, our method acts as one of tfhe steps in the middle-end, and we modify the computational graph if we are solving the UQ problem using the full-grid quadrature points evaluations. 
Below is a coarse outline of how CSDL and our proposed method interact:
\begin{enumerate}
    \item User defines the model and identifies the uncertain inputs.
    \item Generate the quadrature points according to the distributions of the input variables.
    \item Generate the computational graph from the numerical model the user defines.
    \item Modify the computational graph as in Alg.~\ref{alg: ATE}.
    \item Generate an executable back-end.
    \item Post-process the output data to calculate the desired quantities of the output.
\end{enumerate}
    \section{Numerical Results}
\label{Sec: Numerical Results}
We investigate the performance of AMTC by using it with the full-grid NIPC method on three low-dimensional UQ problems involving different computational models. The AMTC method has been implemented in the middle-end of CSDL compiler as a graph transformation method, and all of the computational models involved in the test problems are built in CSDL in order to use AMTC to accelerate its tensor-grid evaluations.

\subsection{Analytical piston model}

 We first consider a UQ problem involving an analytical non-linear model of the cycle time of a piston. This problem is adapted from \cite{ben2007modeling}. The piston cycle time $C$ in seconds is expressed as
\begin{equation}
C=2 \pi \sqrt{\frac{M}{k+S^{2} \frac{P_{0} V_{0} T_{a}}{T_{0} V^{2}}}},
\end{equation}
with
\begin{equation}
V=\frac{S}{2 k}\left(\sqrt{A^{2}+4 k \frac{P_{0} V_{0}}{T_{0}} T_{a}}-A\right) \text { and } A=P_{0} S+19.62 M-\frac{k V_{0}}{S}.
\end{equation}
In this problem, three of the input parameters are normal random variables, and the rest of them are deterministic variables. 
The values and descriptions for all of the variables are shown in Table \ref{tab:piston}.
\begin{table}[h!]
    \centering
    \begin{tabular}{c c c}
         Input parameters & Range & Description  \\
         \hline
         $M$ & $N(50, 10)$ & Piston weight (kg) \\
         $S$ & $N(0.01, 0.002)$ & Piston surface area ($m^{2}$) \\
         $V_{0}$ & $N(0.005, 0.001)$ & Initial gas volume ($m^{3}$) \\
         $k$ & 3000 & Spring coefficient ($N/m$) \\
         $P_{0}$ & 100000 & Atmospheric pressure ($N/m^{2}$) \\
         $T_{a}$ & 293 & Ambient temperature (K) \\
         $T_{0}$ & 350 & Filling gas temperature (K) \\
    \end{tabular}
    \caption{Input parameters and ranges for the piston problem}
    \label{tab:piston}
\end{table}

The objective of this problem is to compute the expectation value of the piston cycle time, namely, $\mathbb{E}[C]$. 

This UQ problem and those that follow are all solved using five methods: integration-based NIPC method using the full-grid quadrature points (full-grid NIPC), full-grid NIPC with the AMTC method (full-grid NIPC with AMTC); integration-based NIPC method using the designed quadrature method (designed quadrature NIPC); kriging and the Monte Carlo method. The kriging method is implemented in its basic form without using adaptive sampling and hyperparameter tuning. The sample points are generated by random sampling, and the kriging surrogate model is trained using the surrogate modelling toolbox in  \cite{SMT2019}.

\begin{figure}%
    \centering
    \subfloat[\centering Comparison of the full-grid NIPC methods with and without applying the AMTC method\label{fig: piston_comp}]{{\includegraphics[width=7cm]{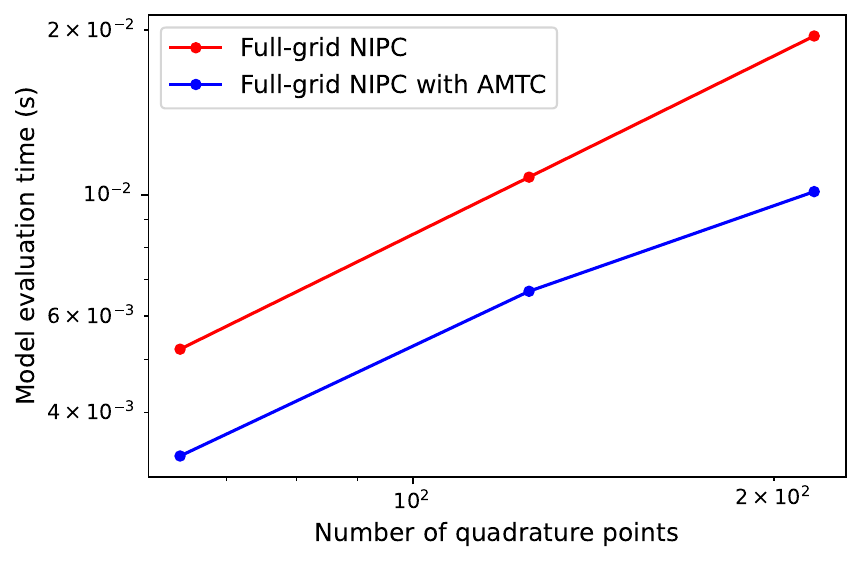} }}%
    \qquad
    \subfloat[\centering Convergence of the UQ results for the five methods\label{fig: piston_results}]{{\includegraphics[width=7cm]{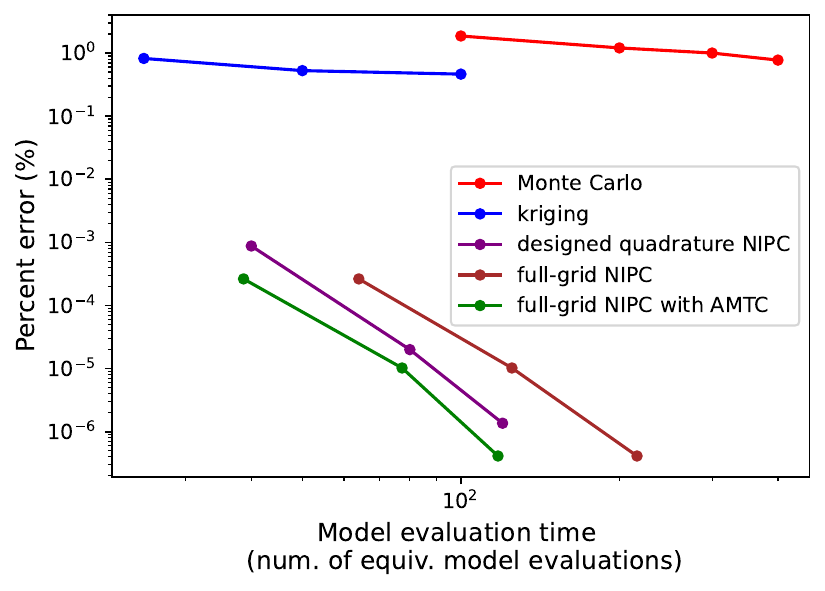} }}%
    \caption{UQ results on the piston model}%
    \label{fig:uq_results_piston}%
\end{figure}

In our numerical experiments, the full-grid NIPC and the full-grid NIPC with AMTC methods always generate the same results. 
Fig.~\ref{fig: piston_comp} shows the model evaluation time in terms of the number of equivalent model evaluations for the full-grid NIPC method with and without AMTC. 
The results show that the AMTC method brought a consistent 50\%-60\% reduction in evaluation time. In Fig.~\ref{fig: piston_results}, the convergence plots of these five methods are compared for the UQ result. 
The percentage errors are calculated with respect to the results of the full-grid NIPC using 400 quadrature points.
For this UQ problem, the NIPC methods completely outperform the Monte Carlo and kriging methods. 
This is because this UQ problem is fairly low-dimensional (only three uncertain inputs), and the function is generally smooth. 
In this kind of UQ problem, PCE-related methods are often the most efficient choices.
Among the NIPC methods, the designed quadrature NIPC performs better than the full-grid NIPC, as the designed quadrature NIPC requires a smaller number of quadrature points to achieve the same level of accuracy on integration. 
However, we see a dramatic speed-up after applying the AMTC method to the full-grid NIPC; the total model evaluation time is reduced by almost an order of magnitude, making the full-grid NIPC with AMTC the most efficient method among other methods.

The major speed-up that AMTC brought for full-grid NIPC can be explained by the sparsity of the influence matrix that describes the dependency information of the computational model in this UQ problem. 
We show the number of dependent operations for each uncertain input in this computational model in Tab. \ref{tab:piston_dep}.
From this table, we observe that out of the 65 operations in the computational graph, 50\% of the operations depend on the uncertain input $M$ and 60\% of the operations depend on uncertain inputs $S$ and $V_0$. 
This means, a large portion of the operations are not dependent on all of the uncertain inputs and a significant number of repeated evaluations are eliminated from the operational level with the AMTC method. 
As this computational model only comprises basic arithmetic operations, the repeated evaluations AMTC eliminated significantly reduces the overall model evaluation time, making the full-grid NIPC the most efficient UQ method.
\begin{table}[h!]
    \centering
    \begin{tabular}{c c c}
         Uncertain input & No. of dependent operations & Total operations  \\
         \hline
         $M$ & 31 & 65 \\
         $S$ & 43 & 65 \\
         $V_{0}$ & 45 & 65 \\
    \end{tabular}
    \caption{Number of dependent operations for each uncertain input in the piston model}
    \label{tab:piston_dep}
\end{table}

\subsection{Low-fidelity multidisciplinary model}
\begin{figure}[hbt!]
\centering
  \includegraphics[width= 7 cm]{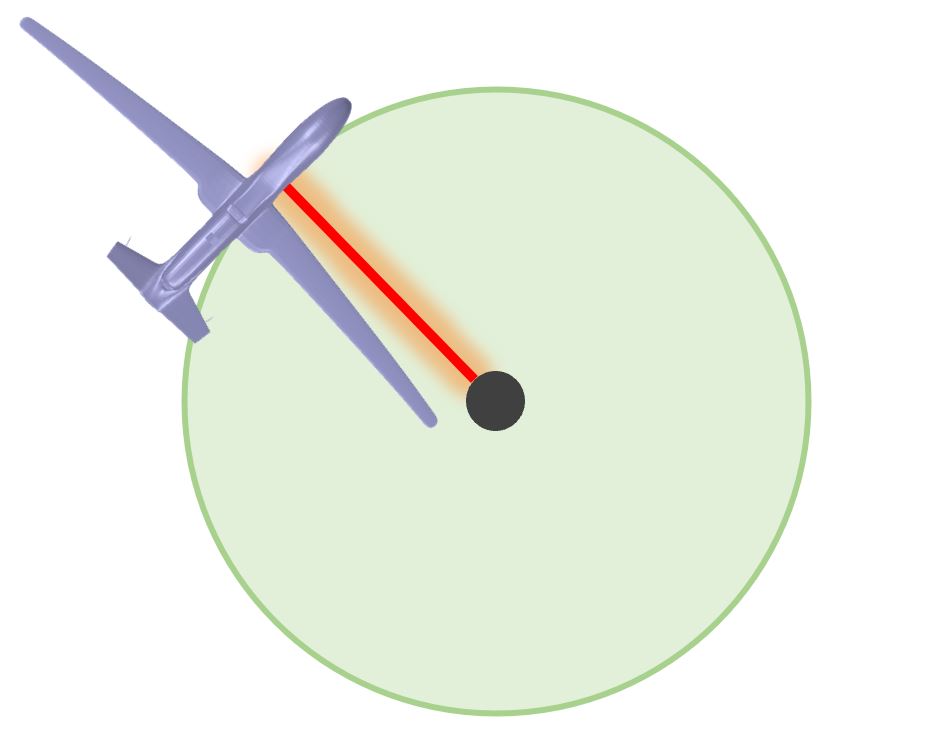}
\caption{A circular cruise mission around a ground station}
\label{fig: mission plot}
\end{figure}

The second UQ problem we consider involves a low-fidelity multidisciplinary model that computes the total energy stored for a laser-beam-powered unmanned aerial vehicle (UAV) cruising around a ground station for one cycle while being charged by the laser beam.
The UQ problem aims to compute the expectation of the total energy stored under three uncertain inputs. We show the mission plot in Fig.~\ref{fig: mission plot} and the input parameters in Tab.~\ref{tab:darpa}.
The computational model involves five disciplines, comprising beam propagation, weights, aerodynamics, and performance models. Further details can be found in \cite{wang2023optimally}.

\begin{figure}%
    \centering
    \subfloat[\centering Comparison of the full-grid NIPC methods with and without applying the AMTC method]{{\includegraphics[width=7cm]{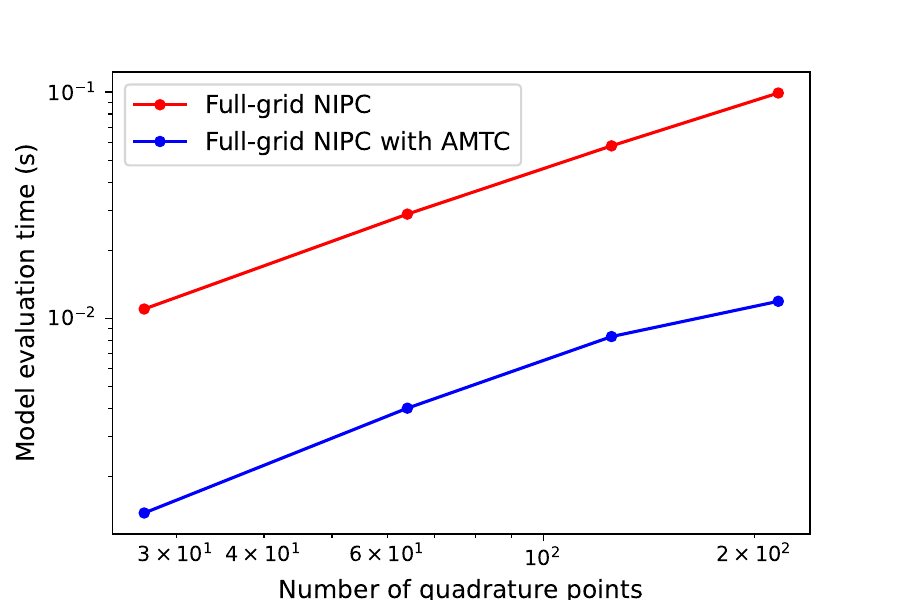} }\label{fig: darpa_comp}}%
    \qquad
    \subfloat[\centering  Convergence of the UQ results for the five methods]{{\includegraphics[width=7cm]{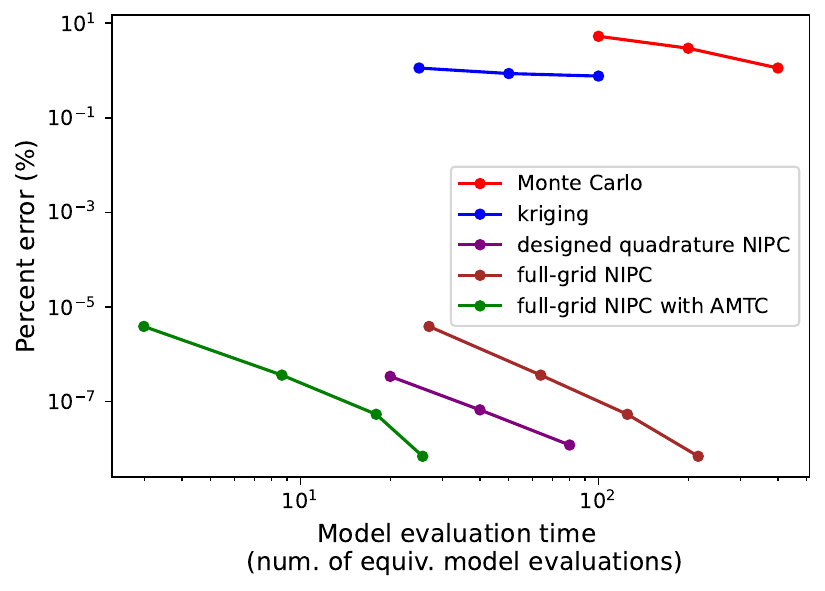} }\label{fig: darpa_results}}%
    \caption{UQ results on the multi-disciplinary model}%
    \label{fig:darpa}%
\end{figure}

The performance comparison of the full-grid NIPC method with and without AMTC is shown in Fig.~\ref{fig: darpa_comp}, and the convergence results for all of the UQ methods are shown in Fig.~\ref{fig: darpa_results}.
We observe that the AMTC method provided approximately a 90\% speed-up in the model evaluation time for the full-grid NIPC method.
As a result, full-grid NIPC with AMTC is more efficient to use than all of the other UQ methods we implemented.
This is not surprising to see, as in the multidisciplinary model, the parameter uncertainties typically come from different disciplines and there may exist some uncertain inputs that only affect a small portion of the operations.
In this scenario, the acceleration provided by the AMTC method can be extremely significant when evaluating on the full-grid quadrature points.
From the dependency information shown in Tab.\ref{tab:darpa_dep}, the uncertain input $\eta$ affects only less than 10\% of the operations in the computational model, while the other two operations affect roughly 50\% of the operations.

\begin{table}[h!]
    \centering
    \begin{tabular}{c c c}
         Input parameters & Range & Description  \\
         \hline
         $V$ & {$N(100, 20)$} & \text {Velocity }(m/s) \\
         $h$ & {$N(10,000, 2,000)$} & \text {Altitude}(m) \\
         $\eta$ & {$N(0.2, 0.03)$} & \text {Atmospheric extinction} \\
    \end{tabular}
    \caption{Input parameters and ranges for the multidisciplinary model}
    \label{tab:darpa}
\end{table}
\begin{table}[h!]
    \centering
    \begin{tabular}{c c c}
         Uncertain input & No. of dependent operations & Total operations  \\
         \hline
         $V$ & 162 & 285 \\
          $h$ & 141 & 285 \\
          $\eta$ & 21 & 285 \\
    \end{tabular}
    \caption{ Number of dependent operations for each uncertain input in the multidisciplinary model}
    \label{tab:darpa_dep}
\end{table}

\subsection{Medium-fidelity multi-point model}
\begin{figure}
    \centering
    \includegraphics[width=7cm]{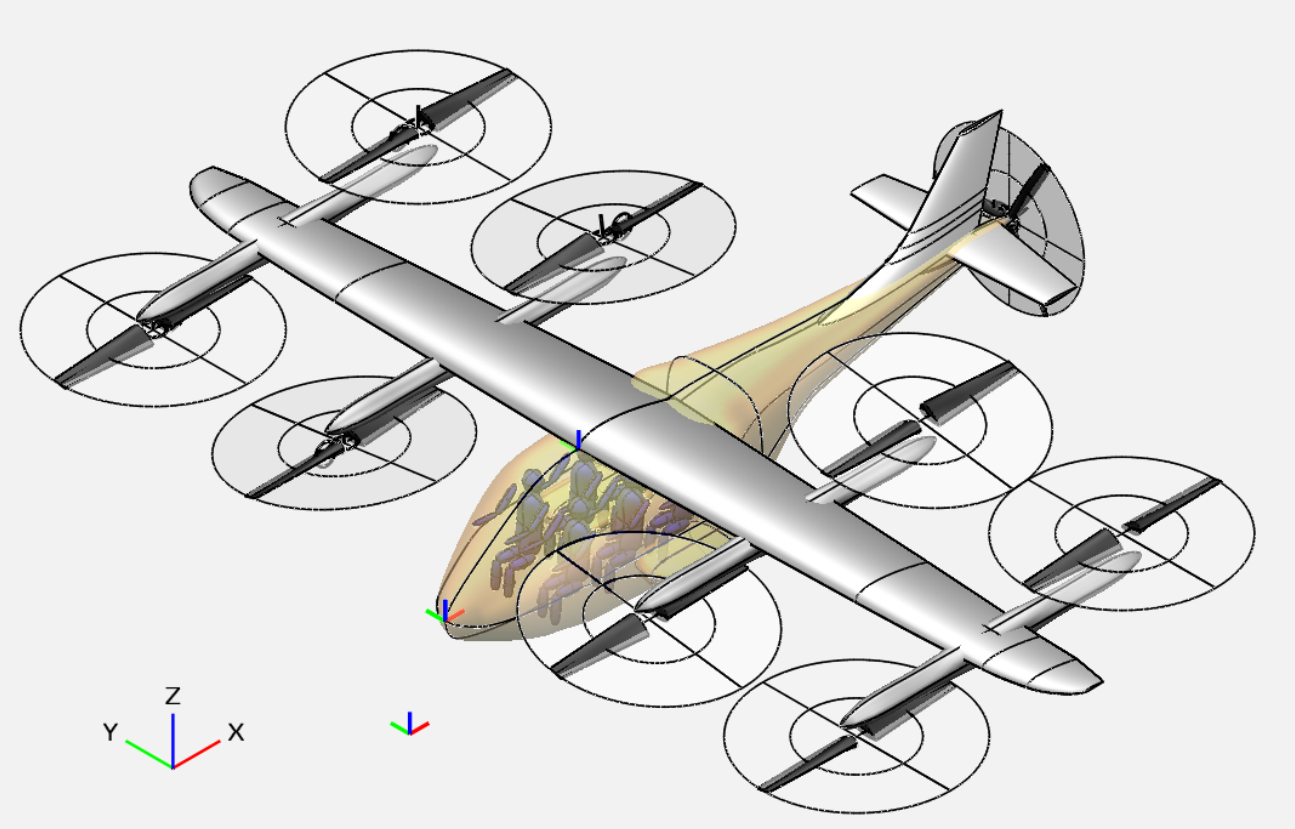}
    \caption{Representation of the eVTOL aircraft~\cite{silva2018vtol}, credit to NASA}
    \label{fig:vtol}
\end{figure}

The third UQ problem we consider is a multi-point mission analysis problem involving an electric vertical takeoff and landing (eVTOL) aircraft. 
The problem is adapted from~\cite{silva2018vtol}. 
The UQ problem aims to compute the expectation of the total energy consumption of a lift-plus-cruise eVTOL aircraft concept (Fig.\ref{fig:vtol}) in a two-segment mission including climb and cruise.
Two uncertain inputs are considered in this problem, which are the flight speeds at two mission segments. The details of the properties for the climb and cruise segment are presented in Table \ref{tab:mission}. For each flight segment, we perform a single-point analysis halfway through the stage and use the vortex-lattice method (VLM), an incompressible and inviscid aerodynamic analysis method to compute the lift and drag forces.
The details of the computational model can be found in \cite{wang2022efficient}.

\begin{table}[h!]
  \begin{center}
    % The table shows results of exact-QN, inexact-QN and adaptive-inexact-QN methods on solving the cantilever bar optimization problem of various problem size. (Opt iterations) is the total number of optimization iterations, and (CG iterations) is the total number of CG iterations.}
    \begin{tabular}{c  c  c } % <-- Alignments: 1st column left, 2nd middle and 3rd right, with vertical lines in between
        & Climb segment
        & Cruise segment
        \\
       \hline
       Initial altitude (ft) & 6,000 & 10,000 \\
       Final altitude (ft) & 10,000 & 10,000 \\
       Flight path angle $\gamma$ (deg) & 20 & 0 \\
       Range $R$ (nmi) & $R_{\text{climb}}$ & $37.5 - R_{\text{climb}}$ \\
       Speed $V$ (Mach) & $N(0.3, 0.03)$ & $N(0.5, 0.05)$ \\
    \end{tabular}
    \caption{
        Properties for climb and cruise segments of the flight mission
    }
    \label{tab:mission}
  \end{center}
\end{table}

\begin{figure}%
    \centering
    \subfloat[\centering Comparison of the full-grid NIPC methods with and without applying the AMTC method]{{\includegraphics[width=7cm]{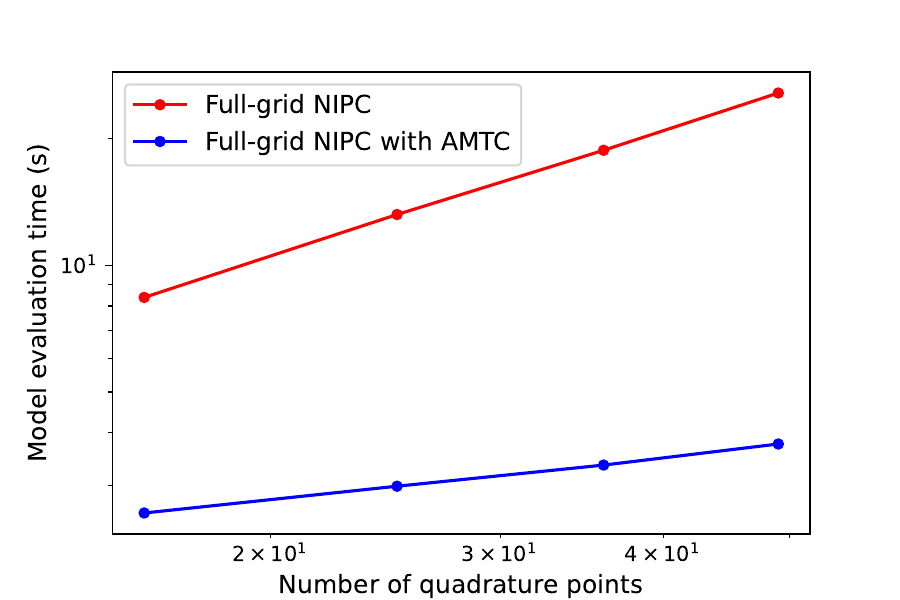}}\label{fig: multi_comp}}%
    \qquad
    \subfloat[\centering  Convergence of the UQ results for the five methods]{{\includegraphics[width=7cm]{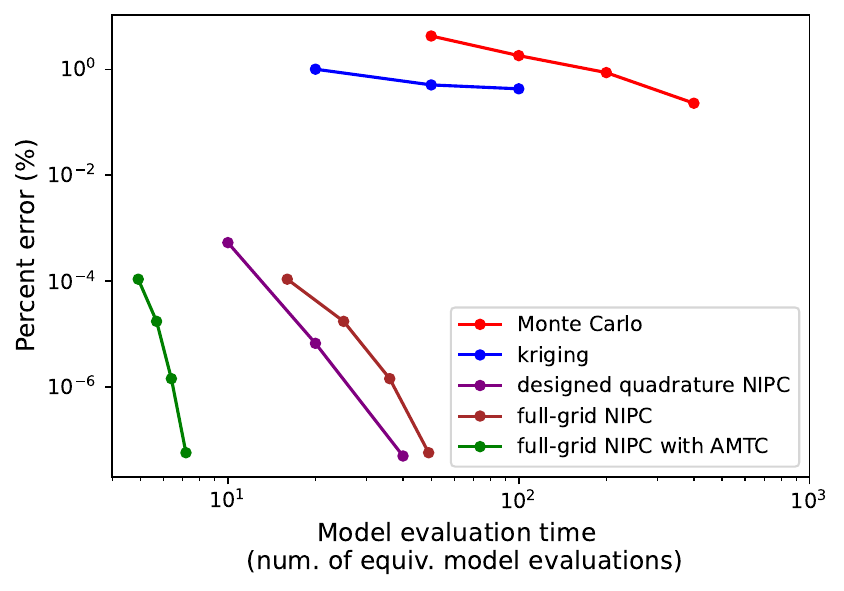} }\label{fig: multi_results}}%
    \caption{UQ results on the multi-point model}%
    \label{fig:multi}%
\end{figure}

We show the performance comparison of the full-grid NIPC method with and without AMTC in Fig.~\ref{fig: multi_comp} and the UQ convergence plot comparing the five UQ methods in Fig.~\ref{fig: multi_results}.
For this problem, the AMTC provided a 70-90\% percent reduction in evaluation time and the reduction got more significant as we increase the number of quadrature points. As a result, the full-grid NIPC with AMTC method is significantly more efficient to use than the other UQ methods. The dependency information in Tab.~\ref{tab:multi_dep} shows roughly 40\% of the operations depend on each uncertain input. The sparsity of the influence matrix is present because, for this multi-point problem, we have two uncertain inputs each only affecting the aerodynamic analysis at one point. 
Although the output of the model is based on the results from both aerodynamic analyses, each aerodynamic analysis is only evaluated on the quadrature points in one-dimensional input space by using the AMTC method.
This dramatically accelerated its model evaluation time on full-grid quadrature points.
In fact, the performance of AMTC can be more significant for many multi-point problems involving multi-point analyses and multiple uncertain inputs.

We recognize that on this test problem, given the straightforward computational graph structure, our method's performance could be replicated by manually conducting two distinct sets of evaluations with the VLM solver and subsequently integrating these evaluations using tensor operations. However, in a practical design workflow, where UQ problems might undergo frequent reformulations, our method can automatically achieve this reduction in computational cost, sparing users the effort of manual implementation.

\begin{table}[h!]
    \centering
    \begin{tabular}{c c c}
         Uncertain input & No. of dependent operations & Total operations  \\
         \hline
         $V_{\text{climb}}$ & 605 & 1505 \\
          $V_{\text{cruise}}$ & 604 & 1505 \\
    \end{tabular}
    \caption{Number of dependent operations for each uncertain input in the multi-point model}
    \label{tab:multi_dep}
\end{table}

\subsection{Blade element momentum rotor model}
The fourth UQ problem we consider is a rotor aerodynamic analysis involving a blade element momentum model. 
The description of the model can be found in~\cite{ruh2023fast}. 
In this UQ problem, we aim to compute the expectation of the total torque generated by the rotor, given two uncertain inputs, rotational rotor speed, and axial free-stream velocity. The description of the uncertain inputs is shown in Tab.~\ref{tab:BEM}.

\begin{figure}%
    \centering
    \subfloat[\centering Comparison of the full-grid NIPC methods with and without applying the AMTC method]{{\includegraphics[width=7cm]{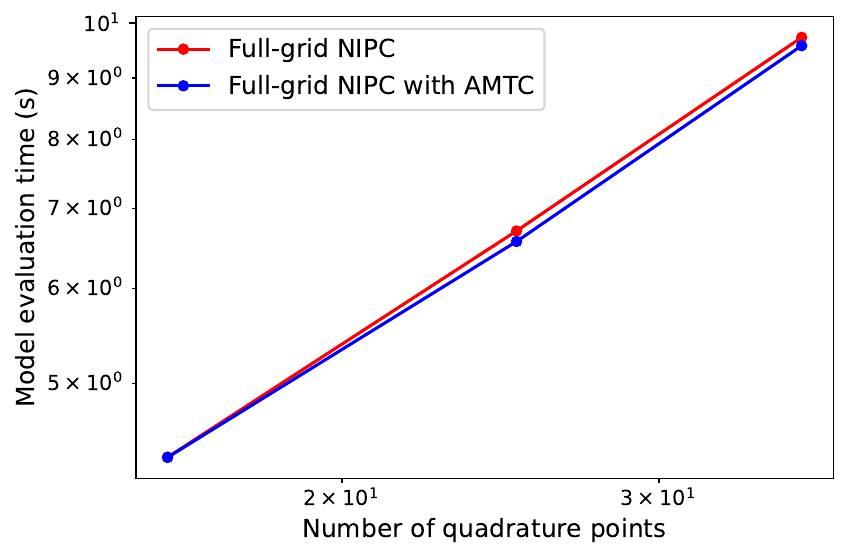}}\label{fig: BEM_comp}}%
    \qquad
    \subfloat[\centering  Convergence of the UQ results for the five methods]{{\includegraphics[width=7cm]{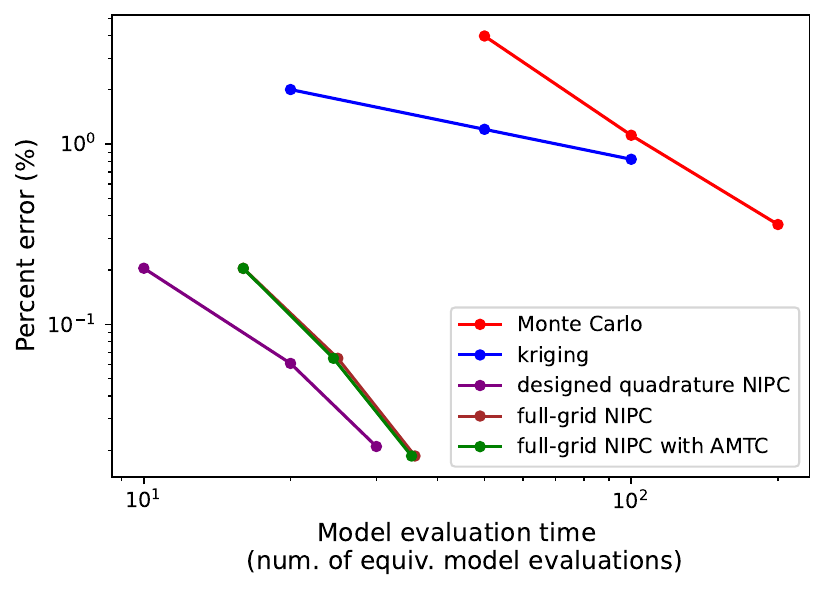} }\label{fig: BEM_results}}%
    \caption{UQ results on the rotor model}%
    \label{fig:BEM}%
\end{figure}

The performance comparison of the full-grid NIPC method with and without AMTC is shown in Fig.~\ref{fig: BEM_comp}. The convergence results for all of the UQ methods are shown in Fig.~\ref{fig: BEM_results}, and the dependency information is shown in Tab.~\ref{tab:BEM_dep}.
In Tab.~\ref{tab:BEM_dep}, we observe that the influence matrix here is also sparse, as roughly 55\% of the operations in the computational model are dependent on each of the uncertain inputs. 
However, we observe little acceleration for the full-grid NIPC using AMTC in Fig.~\ref{fig: BEM_comp}. As a result, the full-grid NIPC is outperformed by the designed quadrature method for the UQ results in Fig.~\ref{fig: BEM_results}.
This is because the dependency information is not the only factor that affects the AMTC method's performance.
The other factor is each operation's evaluation time.
In this model, there exists an implicit operation which also counts as one operation node but its evaluation time takes more than 95\% of the model evaluation time.
Since the AMTC method does not reduce the number of model evaluations on this implicit operation, the reduction in model evaluation time by AMTC becomes insignificant.

\begin{table}[h!]
    \centering
    \begin{tabular}{c c c}
         Input parameters & Range & Description  \\
         \hline
         $\Omega$ & {$N(100, 10)$} & \text {Rotational rotor speed }(rad/s) \\
         $V_x$ & {$N(50, 5)$} & \text {Axial free-stream velocity}(m/s) \\
    \end{tabular}
    \caption{ Input parameters and ranges for the rotor model}
    \label{tab:BEM}
\end{table}

\begin{table}[h!]
    \centering
    \begin{tabular}{c c c}
         Uncertain input & No. of dependent operations & Total operations  \\
         \hline
         $\Omega$ & 398 & 708 \\
          $V_x$ & 423 & 708 \\
    \end{tabular}
    \caption{ Number of dependent operations for each uncertain input in the rotor model}
    \label{tab:BEM_dep}
\end{table}

    \section{Conclusion}
\label{Sec: Conclusion}

This paper introduces a new method, Accelerated Model evaluations on Tensor grids using Computational graph transformations (AMTC), designed to notably reduce the model evaluation cost on tensor grids of input points via computational graph transformation.
It achieves the reduction of model evaluation cost by using the computational graph of the model and algorithmically removing the repeated evaluations on each data node so that it generates the model evaluations on the full-grid input points in an efficient way.
This method can be used with the integration-based non-intrusive polynomial chaos (NIPC) and stochastic collocation (SC) methods to solve forward uncertainty quantification problems.
To generate the numerical results in this paper, we implemented AMTC as a graph transformation method within the three-stage compiler for the Computational System Design Language, so that the modeling language and its compiler automate the process of detecting and capitalizing the model structure, achieving the UQ cost reduction in a general way.
We demonstrated the performance of this method on four test problems, which include an analytical function model, a multidisciplinary model, a multi-point model, and a single-disciplinary model.
For the first three problems, AMTC significantly reduces the model evaluation cost for the full-grid NIPC method, showcasing a speed-up exceeding 50\%. 
%rephrase
Evidently, this positions AMTC as the optimal choice for achieving unparalleled efficiency within UQ methodologies.
However, we did not see this level of performance for the fourth problem as the AMTC does not reduce the number of evaluations on the most cost-dominant operation in the computational graph.
Even though we do not expect this method to achieve a significant reduction of computational cost for every UQ problem, AMTC's application holds promise across a broad range of problems that involve multi-disciplinary models, multi-point models, and other models with sparse computational graph structures. These are the realms in which AMTC is poised to markedly enhance the scalability of NIPC and SC methods.

One limitation of this method is that this method assumes a full-grid structure for the input points.
Looking ahead, a key direction for future work lies in extending AMTC's functionality to encompass partially tensor-structured input points. 
Enabling users to determine the optimal tensor structure based on an assessment of the computational graph structure could lead to even lower model evaluation costs via AMTC.
% promising direction,
Another promising direction is to enable optimization under uncertainty (OUU) with AMTC.
As common OUU frameworks perform a UQ analysis at each optimization iteration, the speedup achieved by AMTC for the UQ analysis is vital to solve certain OUU problems with high-fidelity computational models,
    \section*{Acknowledgments}
The material presented in this paper is, in part, based upon work supported by NASA under award No.~80NSSC21M0070 and DARPA under grant No.~D23AP00028-00.
    \section*{References}

    \printbibliography[heading=none]
    
\end{refsection}

\end{document}